\documentclass[aps,prl,twocolumn,groupedaddress]{revtex4}

\usepackage{graphicx,psfrag}

\newcommand{\bea}{\begin{eqnarray}}
\newcommand{\eea}{\end{eqnarray}}
\newcommand \bra[1]{\left< {#1} \,\right\vert}
\newcommand \ket[1]{\left\vert\, {#1} \, \right>}

\begin{document}


\preprint{TU--713}
\preprint{Mar.\ 2004}

\title{``Coulomb+linear'' form of the static QCD potential
in operator-product-expansion}


\author{Y. Sumino}
\affiliation{Department of Physics, Tohoku University, Sendai, 980-8578 Japan  }


\date{\today}

\begin{abstract}
The static QCD potential is analyzed in operator-product-expansion
within potential-NRQCD framework
when $r \ll \Lambda_{\rm QCD}^{-1}$.
We show that the leading short-distance contribution
to the potential, defined as a perturbatively computable
Wilson coefficient, can be expressed, up to ${\cal O}(r^2)$, 
as a ``Coulomb+linear'' potential.
It coincides with the ``Coulomb+linear'' potential
obtained previously from renormalon-dominance 
hypothesis.
Non-perturbative contributions are ${\cal O}(r^2)$
and subleading.
\end{abstract}


\maketitle

For decades, the static QCD potential $V_{\rm QCD}(r)$ 
has been widely studied 
for the purpose of elucidating the nature of the interaction between heavy
quark and antiquark.
It is
defined from an expectation value of the Wilson loop as
\bea
V_{\rm QCD}(r) 
=
- \! \lim_{T \to \infty} \frac{1}{iT} \,
\ln \frac{\bra{0} {\rm{Tr\, P}} 
e^{ i g \oint_{\cal P} dx^\mu A_\mu }
\ket{0}}
{\bra{0} {\rm Tr} \, {\bf 1} \ket{0}}
,
\eea
where ${\cal P}$ is a rectangular loop of spatial extent $r$ and
time extent $T$.
Generally, $V_{\rm QCD}(r)$ at short-distances can be computed accurately
by perturbative QCD.
On the other hand,
the potential shape at long-distances should be determined by
non-perturbative methods, such as
lattice simulations or phenomenological potential-model analyses
where phenomenological potentials are extracted
from the experimental data for the heavy quarkonium spectra.
Empirically it has been known that phenomenological potentials
and lattice computations of $V_{\rm QCD}(r)$ are both
approximated well by the sum of a Coulomb potential and a linear
potential in the intermediate-distance range.

Since the discovery \cite{Hoang:1998nz} of the 
cancellation of ${\cal O}(\Lambda_{\rm QCD})$
renormalons in the total energy of a static quark-antiquark pair
$E_{\rm tot}(r) \equiv V_{\rm QCD}(r) + 2m_{\rm pole}$,
convergence of the perturbative series for $E_{\rm tot}(r)$
improved drastically and
much more accurate perturbative predictions 
for the potential shape became available.
It was understood that a large uncertainty originating from
the ${\cal O}(\Lambda_{\rm QCD})$
renormalon in $V_{\rm QCD}(r)$ can be absorbed into
twice of the quark pole mass $2m_{\rm pole}$.
Once this is achieved, a perturbative
uncertainty of $E_{\rm tot}(r)$ is 
estimated to be 
${\cal O}(\Lambda_{\rm QCD}^3 r^2)$
at $r \ll \Lambda_{\rm QCD}^{-1}$ \cite{Aglietti:1995tg},
based on the renormalon-dominance hypothesis.

An operator-product-expansion (OPE) 
of $V_{\rm QCD}(r)$ was developed \cite{Brambilla:1999qa} within an 
effective field theory (EFT) 
``potential non-relativistic QCD'' (pNRQCD)
\cite{Pineda:1997bj}.
The idea of OPE is to factorize short-distance contributions
into Wilson coefficients (perturbatively computable)
and non-perturbative contributions into matrix elements of operators,
when the following hierarchy of scales exists:
\bea
\Lambda_{\rm QCD} \ll \mu_f \ll 1/r .
\label{hierarchy}
\eea
Here, $\mu_f$ denotes the factorization scale.
In this framework, residual renormalons, starting from 
${\cal O}(\Lambda_{\rm QCD}^3 r^2)$,
are absorbed into
the matrix elements of non-local operators
(non-local gluon condensates).
Then, in the multipole expansion at $r \ll \Lambda_{\rm QCD}^{-1}$, 
the leading non-perturbative 
contribution to the potential 
becomes ${\cal O}(\Lambda_{\rm QCD}^3 r^2)$ \cite{Brambilla:1999qa}.

Subsequently, several studies 
\cite{Sumino:2001eh,Recksiegel:2001xq}
showed that perturbative
predictions for  $V_{\rm QCD}(r)$ agree well
with phonomenological potentials  
and lattice calculations of $V_{\rm QCD}(r)$, 
once the ${\cal O}(\Lambda_{\rm QCD})$ renormalon contained 
in $V_{\rm QCD}(r)$ is cancelled.
Ref.\ \cite{Lee:2002sn} showed that 
a Borel resummation of the perturbative series gives a potential shape
which agrees with lattice results, if the ${\cal O}(\Lambda_{\rm QCD})$ 
renormalon is properly taken into account.
In fact these agreements hold within
the ${\cal O}(\Lambda_{\rm QCD}^3 r^2)$ uncertainty.
These observations support the validity of renormalon dominance
and of OPE for $V_{\rm QCD}(r)$.

Once the ${\cal O}(\Lambda_{\rm QCD})$ renormalon is cancelled,
the perturbative QCD potential becomes steeper than
the Coulomb potential as $r$ increases.
This feature is understood, within perturbative QCD, 
as an effect of the {\it running} of the strong coupling constant 
\cite{Brambilla:2001fw,Sumino:2001eh}.

Moreover,
using a scale-fixing prescription based on renormalon dominance hypothesis,
it was shown analytically \cite{Sumino:2003yp} that the
perturbative QCD potential approaches a ``Coulomb+linear''
form at large orders, up to an ${\cal O}(\Lambda_{\rm QCD}^3 r^2)$ uncertainty.
The ``Coulomb+linear'' potential can be computed systematically 
as more terms of perturbative series are included
via renormalization-group (RG); up to the next-to-next-to-leading 
logarithmic order (NNLL),
it shows a convergence towards lattice results.

In this paper, we analyze the QCD potential using OPE developed
in \cite{Brambilla:1999qa} and compare the leading 
Wilson coefficient (singlet potential)
with the ``Coulomb+linear'' potential obtained 
in \cite{Sumino:2003yp}.

The 
$V$-scheme coupling in momentum space $\alpha_V(q)$ is defined as
\bea
V_{\rm QCD}(r) &=&
\int \frac{d^3\vec{q}}{(2\pi)^3} \, e^{i \vec{q} \cdot \vec{r}}
\, \biggl[
-4 \pi  C_F \, \frac{\alpha_V(q)}{q^2}
\biggr]
\\
&=&
- \frac{2C_F}{\pi}
\int_0^\infty dq \, \frac{\sin (qr)}{qr} \, \alpha_V(q) ,
\label{one-param-int}
\eea
where $q=|\vec{q}|$;
$C_F$ is the second Casimir operator of the fundamental 
representation.
In perturbative QCD, $\alpha_V(q)$ is calculable in a series
expansion of the strong coupling constant:
\bea
\alpha_V^{\rm PT}(q) 
&=& \alpha_S \, \sum_{n=0}^{\infty} P_n(\ln (\mu/q) ) \,
\biggl( \frac{\alpha_S}{4\pi} \biggr)^n
\\
&=& 
\alpha_S(q) \, \sum_{n=0}^{\infty} a_n \,
\biggl( \frac{\alpha_S(q)}{4\pi} \biggr)^n 
,
\label{alfV}
\eea
where,
$P_n(\ell)$ denotes an $n$-th-degree polynomial of $\ell$
and $a_n = P_n(0) $.
In this paper, unless the argument is specified explicitly,
$\alpha_S \equiv \alpha_S(\mu)$ denotes the strong coupling constant
renormalized at the renormalization scale $\mu$,
defined in the $\overline{\rm MS}$ scheme.
Here and hereafter,
$\alpha_V^{\rm PT}(q)$ represents a perturbative evaluation of 
$\alpha_V(q)$ supplemented by RG evolution of $\alpha_S(q)$.
For instance, by $\alpha_V^{\rm PT}(q)$
up to NNLL, we mean that in (\ref{alfV}) the sum is taken for
$n \leq 2$ and the three-loop running coupling is used for $\alpha_S(q)$.

The ``Coulomb+linear'' potential $V_{\rm C+L}(r)$
obtained in \cite{Sumino:2003yp}, up to NNLL, is given by
\bea
&&
V_{\rm C+L}(r) = V_{\rm C}(r) + \sigma \, r, 
\label{CplusLpot}
\\
&&
V_{\rm C}(r) = - \frac{4\pi C_F}{\beta_0 r} 
- \frac{2C_F}{\pi} \, {\rm Im}
\int_{C_1}\! dq \, \frac{e^{iqr}}{qr} \, \alpha_V^{\rm PT}(q) 
,
\label{Vc}
\\ &&
{\sigma} = \frac{C_F}{2\pi i} \int_{C_2}\! dq \, q \, \alpha_V^{\rm PT}(q) ,
\label{sigma}
\eea
where $\beta_n$ represents the $(n+1)$-loop coefficient of the
beta function; e.g.\ in $SU(3)$ Yang-Mills theory, 
$\beta_0=11$, $\beta_1=102, ~\cdots$.
The integral paths $C_1$ and $C_2$ are displayed 
in Fig.~2 of \cite{Sumino:2003yp}.
The coefficient of the linear potential $\sigma$ 
can be expressed
analytically in terms of the Lambda parameter in the 
$\overline{\rm MS}$-scheme
$\Lambda_{\overline{\rm MS}}$.
The ``Coulomb'' potential has a short-distance
asymptotic behavior consistent with RG,
$V_{\rm C}(r) \sim -2\pi C_F (\beta_0r)^{-1} \Bigl[
\ln\Bigl(\frac{1}{r \Lambda_{\overline{\rm MS}}}\Bigr) + 
\frac{\beta_1}{2\beta_0^2}\ln\ln\Bigl(\frac{1}{ r \Lambda_{\overline{\rm MS}}}
\Bigr)\Bigr]^{-1}$,
whereas its long-distance behavior is given by
$V_{\rm C}(r) \sim -4\pi C_F /(\beta_0r)$;
in the intermediate region both asymptotic forms are smoothly
interpolated.

Let us first present an intuitive argument.
In fact, it already embraces an essential part of our discussion.
We separate the integral (\ref{one-param-int}) into the 
regions $q>\mu_f$ and $q<\mu_f$.
In the former region, 
$\alpha_V(q)$ can be approximated well by $\alpha_V^{\rm PT}(q)$, hence
we define
\bea
V_{\rm UV}(r;\mu_f) = - \frac{2C_F}{\pi}
\int_{\mu_f}^\infty dq \, \frac{\sin (qr)}{qr} \, \alpha_V^{\rm PT}(q) .
\eea
Since $\mu_f \gg \Lambda_{\rm QCD}$, 
we expect that an accurate perturbative prediction for $V_{\rm UV}(r;\mu_f)$
can be made.
In the region $q<\mu_f$,
$\alpha_V(q)$ cannot be evaluated reliably in perturbation theory;
rather it should be determined
non-perturbatively.
On the other hand, we may expand in $r$ since $\mu_f r \ll 1$
by eq.\,(\ref{hierarchy}):
\bea
V_{\rm IR}(r;\mu_f) &=& - \frac{2C_F}{\pi}
\int_0^{\mu_f}\!\! dq \, \biggl[ \, 1 - \frac{q^2r^2}{6}  + \dots
\biggr] \, \alpha_V(q) 
\nonumber \\
&=&
{\rm const.} + {\cal O}(\mu_f^3r^2) .
\label{VIR}
\eea
We can show that
\bea
V_{\rm UV}(r;\mu_f) - V_{\rm C+L}(r) = 
{\rm const.} + {\cal O}(\mu_f^3r^2) .
\label{essence}
\eea
Eqs.\,(\ref{VIR}) and (\ref{essence}) imply that the 
``Coulomb+linear'' part of the QCD potential is determined by
the short-distance contributions ($q>\mu_f$), hence it is predictable
in perturbative QCD,
while the non-perturbative contributions are of order
$\mu_f^3r^2$ and subleading at $r \ll \mu_f^{-1}$.
[Throughout this paper, we are not concerned about 
the constant part of $V_{\rm QCD}(r)$, keeping in mind
that it can always be absorbed into $2m_{\rm pole}$ in the
total energy $E_{\rm tot}(r)$.]

\begin{figure}
\psfrag{C3}{$C_3$} 
\psfrag{q}{$q$} 
\psfrag{muf}{$\mu_f$} 
\psfrag{q*}{$q_*$} 
\psfrag{0}{\hspace{-1mm}\raise-1mm\hbox{$0$}}
\includegraphics[width=4.5cm]{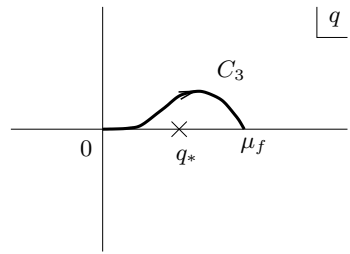}
\caption{
Integral path $C_3$ in the complex $q$-plane.
$q_*$ denotes the IR singularity of $\alpha_S(q)$.
For 1-loop running, $q_*$ is a pole; beyond 1-loop running,
$q_*$ is a branch point.
In the latter case, branch cut is on the real axis starting from $q_*$
to $-\infty$.
\label{path}}
\end{figure}

Eq.~(\ref{essence}) can be shown as follows.
According to (\ref{Vc}),
\bea
&&
V_{\rm UV}(r;\mu_f) - V_{\rm C+L}(r) 
\nonumber\\
&&
= 
\frac{4\pi C_F}{\beta_0 r} + \frac{2C_F}{\pi} \, {\rm Im}
\int_{C_3}\! dq \, \frac{e^{iqr}}{qr} \, \alpha_V^{\rm PT}(q) 
-\sigma r ,
\label{proof}
\eea
where the integral path $C_3$ is shown in Fig.~\ref{path}.
Since $\mu_f r \ll 1$, we may expand the Fourier factor as
$e^{iqr} = 1 + iqr - \frac{1}{2}(qr)^2 + \dots$
in the above integral.
Then one can show by suitable change of variables 
that the $r^{-1}$ term
\bea
\frac{2C_F}{\pi} \, {\rm Im}
\int_{C_3}\! dq \, \frac{\alpha_V^{\rm PT}(q) }{qr} 
= - \frac{C_F}{\pi i} 
\int_{C_2}\! dq \, \frac{\alpha_V^{\rm PT}(q) }{qr} 
\eea
equals $-4\pi C_F/ (\beta_0r)$ (at least) up to
NNLL.
Similarly, one can show for the $r^{1}$ term
\bea
\frac{2C_F}{\pi} \, {\rm Im}
\int_{C_3}\! dq \, \Bigl( -\frac{1}{2} {qr} \Bigr)  \, {\alpha_V^{\rm PT}(q) }
= \sigma \, r .
\label{proof3}
\eea
Therefore, only remaining terms on the right-hand-side of
(\ref{proof}) are 
${\rm const.} + {\cal O}(\mu_f^3r^2) $.

Being intuitive, the above argument is subject to some flaws:
(i) A factorization of scales is introduced only in the
integral over $q$, whereas in a consistent OPE one should 
factorize scales in all quantum effects, namely in the computation
of $\alpha_V(q)$ as well.
(ii) It is known that $a_n$ ($n\geq 3$) in
$\alpha_V^{\rm PT}(q)$ includes
IR divergences \cite{Appelquist:tw}, 
so that $V_{\rm UV}(r;\mu_f)$
is not well-defined beyond NNLL.
(iii) The perturbative series of $V_{\rm UV}(r;\mu_f)$ may still be an
asymptotic series, hence one should clarify
how to define $V_{\rm UV}(r;\mu_f)$.
All these points (i)--(iii) are remedied in a consistent framework
of OPE, in dimensional regularization and with appropriate
renormalization procedure.

An OPE of $V_{\rm QCD}(r)$ was developed in \cite{Brambilla:1999qa}.
In this and the next paragraph, we review the content of 
that paper relevant to our analysis.
Within this framework, short-distance contributions are contained
in the potentials, which are in fact the Wilson coefficients,
while non-perturbative contributions are contained in the
matrix elements that are organized 
in multipole expansion in $\vec{r}$ at $r \ll \Lambda_{\rm QCD}^{-1}$.
The following relation was derived: 
\bea
&&
V_{\rm QCD}(r) = V_S(r) + \delta E_{\rm US}(r),
\label{OPE}
\\&&
\delta E_{\rm US}=
- i g^2 \frac{T_F}{N_C}
\int_0^\infty \! \! dt \, e^{-i \, \Delta V(r)\, t} 
\nonumber\\&&
~~~~~~~~~~~~~~~~~~
\times
\langle \vec{r}\cdot\vec{E}^a(t) \varphi_{\rm adj}(t,0)^{ab}
\vec{r}\cdot\vec{E}^b(0) \rangle
\nonumber\\&&
~~~~~~~~~~~~~~~~~~
+{\cal O}(r^3) .
\label{deltaEUS}
\eea
$V_S(r)$ denotes the singlet potential.
$\delta E_{\rm US}(r)$ denotes the non-perturbative 
contribution to the QCD potential, which starts at 
${\cal O}(\Lambda_{\rm QCD}^3 r^2)$ in the multipole expansion.
$\Delta V(r) = V_O(r) - V_S(r)$ denotes the
difference between the octet and singlet potentials;
see \cite{Brambilla:1999qa} for details.
Intuitively $V_S(r)$ corresponds to $V_{\rm UV}(r;\mu_f)$
and $\delta E_{\rm US}(r)$ to $V_{\rm IR}(r;\mu_f)$.
We adopt dimensional regularization in our analysis;
we also refer to hard cutoff schemes when discussing conceptual aspects.

In perturbative expansion in $\alpha_S$, the QCD potential 
$V_{\rm QCD}(r)$ coincides
with the singlet potential $V_S(r)$, i.e.\ $\delta E_{\rm US}=0$.
As already mentioned, perturbative expansion of
$V_{\rm QCD}(r)$ in $\alpha_S$
includes IR divergences beyond ${\cal O}(\alpha_S^3)$, hence
$V_S(r)$ also includes IR divergences in dimensional regularization.
$\delta E_{\rm US}(r)$ is expected to be non-zero beyond perturbation
theory.
In fact, if we do not expand $\Delta V(r)$ in $\alpha_S$ in
(\ref{deltaEUS}) \footnote{
This is consistent with the concept of the EFT, since this theory is
assumed to correctly describe physics at scales much below $1/r$. 
In this case $\Delta V(r)$ ($ \ll 1/r$) should be kept in the denominator
of the propagator $[E-\Delta V(r)]^{-1}$.
}
(but expand all other factors), $\delta E_{\rm US}(r)$ becomes non-zero
since $\Delta V(r)$ acts as an IR regulator.
In this case, $\delta E_{\rm US}(r)$ contains
UV divergences, given as poles in $\epsilon$, which exactly cancel 
the poles corresponding to the IR divergences in $V_S(r)$.
Consequently, in the sum (\ref{OPE}), $V_{\rm QCD}(r)$ becomes
finite as $\epsilon \to 0$.
These divergences in $V_S(r)$ and $\delta E_{\rm US}(r)$, respectively,
can be regarded as artefacts of dimensional regularization, where the
integral regions of virtual momenta extend from 0 to $\infty$.
If we introduce a hard cutoff to each momentum integration, corresponding
to the factorization scale $\mu_f$, 
$V_S(r)$ ($q>\mu_f$) and $\delta E_{\rm US}(r)$ ($q<\mu_f$),
respectively, would become finite and dependent on $\mu_f$.
In dimensional regularization, 
$V_S(r)$ can be made finite
by multiplicative renormalization,
i.e.\ by adding
a counter term $(Z_S-1) V_S(r)$.

\begin{figure}
\psfrag{singlet}{singlet} 
\psfrag{k}{$k$} 
\psfrag{octet}{octet} 
\includegraphics[width=4.5cm]{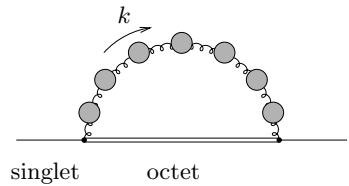}
\caption{
Graph for $\delta E_{\rm US}(r)$ at ${\cal O}(r^2)$
in the large-$\beta_0$ approximation.
Bubble chain represents the renormalized
gluon propagator in the large-$\beta_0$ approximation,
$\displaystyle
\frac{i}{k^2 [1-\Pi(k^2)]}$ with
$\displaystyle
\Pi (k^2) = \frac{\beta_0\alpha_S}{4\pi}
\biggl[ \Bigl(\frac{e^{\gamma_E}\mu^2}{-k^2}\Bigr)^\epsilon
\times
\frac{6\Gamma(\epsilon)\Gamma(2-\epsilon)^2}{\Gamma(4-2\epsilon)}
-\frac{1}{\epsilon} \biggr]
$.
\label{large-beta0}}
\end{figure}

With respect to the spirit of factorization in OPE,
it is natural to subtract IR renormalons from $V_S(r)$ in a similar manner.
In \cite{Pineda:2001zq}, this was advocated and
in practice subtraction of the 
${\cal O}(\Lambda_{\rm QCD})$ renormalon was carried out explicitly.
The known IR renormalons of $V_S(r)$
[$=$perturbative expansion of $V_{\rm QCD}(r)$]
are contained in the integral
\cite{Beneke:1998ui}\footnote{
Here, we neglect the contributions of the instanton-induced singularities
\cite{Beneke:1998ui}
on the positive real axis in the Borel
plane.
These contributions are known to be rather small in any case.
}
\bea
&&
\int_0^{\mu_f} \! dq \, \frac{\sin (qr)}{qr} \, \alpha_V^{\rm PT}(q) 
=
\int_0^{\mu_f} \! dq \,
\biggl( { 1 \! - \! \frac{q^2r^2}{6} \!+\! \cdots} \biggr)
\nonumber\\
&&~~~~~~~~~~~~~~~~
\times
\biggl[ \alpha_S(q) + a_1 \Bigl( \frac{\alpha_S(q)}{4\pi} \Bigr)^2
+ \cdots \biggr]
.
\label{int-renormalon}
\eea
In a hard cutoff scheme, it was shown \cite{Brambilla:1999qa} that the
${\cal O}(\Lambda_{\rm QCD}^3r^2)$ IR renormalon of $V_S(r)$ 
can be absorbed into $\delta E_{\rm US}(r)$.
In dimensional regularization ($D = 4-2\epsilon$), 
one may compute, for instance, 
$\delta E_{\rm US}(r)$ at ${\cal O}(r^2)$
in the large-$\beta_0$ approximation \cite{Beneke:1994qe}, correponding to the
graph in Fig.~\ref{large-beta0}.
It is given by
\bea
&&
\delta E_{\rm US}(r)\Bigr|_{\mbox{\scriptsize{large-}}\beta_0}
= \frac{C_F\alpha_S}{4\pi} \times
8 r^2 \Delta V(r)^3
\nonumber\\
&&
~~~~~~~~~~
\times \sum_{n=0}^\infty
\biggl( \frac{\beta_0\alpha_S}{4\pi} \biggr)^n
\biggl[ \, n! \, G_{n+1} + \frac{1}{\epsilon^{n+1}}\frac{(-1)^n}{n+1}
g(\epsilon) 
\biggr]
\nonumber 
\\ &&
~~~~~~~~~~~~~~~~~~~
+ {\cal O}(\epsilon , \, r^3) ,
\\ &&
G(u) \equiv \sum_{j=0}^\infty G_j \, u^j 
\nonumber\\
&&
~~~~~~
=
\Biggl[
\frac{\mu \, e^{5/6}}{2\,\Delta V(r)} \Biggr]^{2u} 
\frac{2\, \Gamma (2-u) \Gamma (2u-3)}{\Gamma (u-1)} ,
\\ &&
g(\epsilon) = \frac{\Gamma (4-2\epsilon)}{36\,
\Gamma (1+\epsilon) \Gamma(2-\epsilon)^2 \Gamma(1-\epsilon)} .
\eea
We note that the renormalon contribution ($n! \, G_{n+1}$) 
and the
UV divergences (multiple poles in $\epsilon$) 
are included in separate parts in this approximation.
One can check explicitly that
the ${\cal O}(\Lambda_{\rm QCD}^3r^2)$ UV renormalon, corresponding to the
pole at $u=3/2$ of $G(u)$, cancels the ${\cal O}(\Lambda_{\rm QCD}^3r^2)$
IR renormalon in $V_S(r)$ 
\cite{Aglietti:1995tg} in the large-$\beta_0$ approximation.
Therefore, in dimensional regularization,
it is appropriate to subtract from $V_S(r)$ the IR renormalons,
e.g.\ in the form of
(\ref{int-renormalon}).

We define a renormalized singlet potential (in dimensional regularization),
in a scheme where the IR divergences and IR renormalons are subtracted, as
\bea
V_S^{({\rm R})}\!(r;\mu_f\!) = 
 - \frac{2C_F}{\pi} \!\!
\int_{\mu_f}^\infty \!\!\!\! dq \, \frac{\sin (qr)}{qr} \, 
[ \alpha_V^{\rm PT}\! (q) + \delta\alpha_V\! (q) ] .
\label{VSR}
\eea
$\delta\alpha_V(q)$ is the counter term which subtracts the
IR divergences, given as multiple poles in $\epsilon$,
e.g.\ in the $\overline{\rm MS}$ scheme.\footnote{
In principle, one should compute $D$-dimensional Fourier integral of
$\alpha_V(q)$ defined in $D$ dimensions,
subtract the IR divergences, and then take the limit $\epsilon \to 0$.
In our case, it coincides with the naive expression (\ref{VSR}).
}
The dependence on $\mu_f$ is introduced through
subtraction of the IR divergences and IR renormalons.
Then, we can apply our argument given through
eqs.~(\ref{proof})--(\ref{proof3}) to show that
\bea
V_S^{({\rm R})}(r;\mu_f) - V_{\rm C+L}(r) = 
{\rm const.} + {\cal O}(\mu_f^3r^2) ,
\label{result}
\eea
up to NNLL (since $\delta\alpha_V(q)$ can be taken as zero
up to this order).
Moreover, beyond NNLL, this relation
still holds after a simple replacement
$\alpha_V^{\rm PT}(q) \to \alpha_V^{\rm PT}(q) + \delta\alpha_V(q)$
in the definition of $V_{\rm C+L}(r)$, (\ref{Vc}) and (\ref{sigma}).

One may think that subtracting the integral 
(\ref{int-renormalon}) is not sufficient for 
subtracting all the IR renormalons.
Our result (\ref{result}) is unchanged, even
if one subtracts the IR renormalon contributions
using whatever other sophisticated method for estimating them.
This is because the IR renormalons in $V_S(r)$ take the form
${\rm const.} + {\cal O}(\Lambda_{\rm QCD}^3r^2)$.

The perturbative expansion of $V_S^{({\rm R})}(r;\mu_f)$ 
may still be an asymptotic series.
Since the IR renormalons have been subtracted and the factorization
scale is set as $\mu_f \gg \Lambda_{\rm QCD}$, we may expect that 
$V_S^{({\rm R})}(r;\mu_f)$ is Borel summable.\footnote{
This is up to the uncertainties caused by the instanton-induced singularities
\cite{Beneke:1998ui} in the Borel plane.
}
(At least, the Borel integral is convergent 
in the large-$\beta_0$ approximation.)
Then, we may define $V_S^{({\rm R})}(r)$ from the perturbative series
either by Borel summation or according to the prescription of
\cite{Beneke:1992ea};
both prescriptions lead to the same result when the series is Borel summable.

Our result (\ref{result}) shows that the renormalized singlet
potential $V_S^{({\rm R})}(r)$ can be expressed as a 
``Coulomb+linear'' potential $V_{\rm C+L}(r)$, 
up to ${\cal O}(\mu_f^3r^2)$, at short distances.
We re-emphasize that there is no
freedom to add a linear potential to $V_{\rm C+L}(r)$
in (\ref{result}) \cite{Brambilla:1999qa}.

On the other hand, there is an arbitrariness 
in how to separate $V_{\rm C+L}(r)$
into ``Coulomb'' and linear parts, as discussed in \cite{Sumino:2003yp}.
Stating more accurately, as yet
we do not know any mathematically well-defined principle
to separate $V_{\rm C+L}(r)$ into Coulombic and linear parts
about $r \sim 0$, because of
$1/\ln r$ dependence in $V_{\rm C+L}(r)$.
Nonetheless, we consider
the present separation (\ref{CplusLpot})--(\ref{sigma}) 
a natural one according to its construction, and also because
it is demonstrated 
that the perturbative prediction of
$V_{\rm QCD}(r)$ up to ${\cal O}(\alpha_S^N)$ is 
approximated well by this ``Coulomb+linear'' form
for a fairly wide range of $r$ and $N$ \cite{Sumino:2003yp}.

\begin{acknowledgments}
The author is grateful to A.~Pineda and M.~Tanabashi
for enlightening discussion.
\end{acknowledgments}


\end{document}